%% file: ICDM/main.tex
\documentclass[conference]{IEEEtran}

\usepackage{eso-pic}
\AddToShipoutPictureFG*{%
  \AtPageLowerLeft{%
    \put(0,36){\makebox[\paperwidth][c]{\footnotesize
      \copyright~2026 IEEE. Personal use of this material is permitted. Permission from IEEE must be obtained for all other uses.}}%
  }%
}
\usepackage{cite}
\usepackage{amsmath,amssymb,amsfonts}
\usepackage{algorithmic}
\usepackage{graphicx}
\usepackage{textcomp}
\usepackage{xcolor}
\usepackage{booktabs}
\usepackage[caption=false,font=footnotesize]{subfig}
\usepackage{tabularx}
\usepackage{longtable}
\usepackage{multirow}

\def\BibTeX{{\rm B\kern-.05em{\sc i\kern-.025em b}\kern-.08em
    T\kern-.1667em\lower.7ex\hbox{E}\kern-.125emX}}

\begin{document}

\title{Comprehensive Workflow for Uncertainty-Aware Ionic Liquid
Solvent Discovery\\

}

\author{
  \IEEEauthorblockN{Alexander Lobo}
  \IEEEauthorblockA{
    \textit{BCG X AI Science Institute}\\
    Boston, USA \\
    lobo.alexander@bcg.com}
  \and
  \IEEEauthorblockN{Urvi Awasthi}
  \IEEEauthorblockA{
    \textit{BCG X AI Science Institute}\\
    New York, USA \\
    awasthi.urvi@bcg.com}
    \and
  \IEEEauthorblockN{Leonid Zhukov}
  \IEEEauthorblockA{
    \textit{BCG X AI Science Institute}\\
    Dubai, UAE \\
    zhukov.leonid@bcg.com}

}

\maketitle

\begin{abstract}
Designing synthesizable materials with targeted properties remains one of the central objectives in materials sciences.  Ionic liquids (ILs) are well suited to this goal due their “tunability”, where physical and chemical properties can be precisely customized by modifying their molecular structure, specifically by varying the combination of cations and anions. At the same time, current data-driven property models for ionic liquids often degrade in scarce, noisy, and uneven data regimes, where measurements cover only a small fraction of chemical space and a limited range of thermodynamic conditions. This makes out-of-distribution screening brittle: even when point predictions look plausible, it is difficult to compare candidates consistently, quantify downstream implications, or trust rankings enough to guide molecular discovery. We present a property-prediction and optimal candidate selection workflow that treats data processing, deep learning modeling, and physics-based calculations as a coupled system, linking property prediction with uncertainty estimation and physics-based filtering. By tightening the link between predicted properties and the criteria used for down-selection, the workflow supports reliable out-of-distribution ranking and enables prioritization of previously untested ionic liquid candidates with favorable predicted trade-offs. We demonstrate the application of our workflow by designing novel ionic liquids for CO$_2$ capture from flue gas, predicting their properties and selecting Pareto-optimal candidates based on CO$_2$ solubility, IL viscosity and working capacity.
\end{abstract}


\input{ICDM/sections/00_intro}

\input{ICDM/sections/01_background}

\input{ICDM/sections/02_methods}

\input{ICDM/sections/03_results}

\input{ICDM/sections/99_conclusions}

\input{ICDM/sections/04_limitations}

\input{ICDM/sections/genai}

\bibliographystyle{IEEEtran}
\bibliography{bibliography}

\input{ICDM/sections/999_appendix}

\end{document}

%% file: ICDM/sections/00_intro.tex
\section{Introduction}

Designing synthesizable materials with targeted properties is central to materials science. Ionic liquids are attractive process solvents because their properties can be systematically tailored by modifying the cation--anion combination, earning them the moniker ``designer solvents'' and enabling large combinatorial design spaces \cite{miao2014chemical,borun2019conductance,ghandi2014review}.

However, the experimental record for IL thermophysical properties is sparse and uneven, covering only a small fraction of chemical space and limited thermodynamic conditions, with noisy or inconsistent reported values. This makes out-of-distribution screening brittle: point predictions may appear plausible, yet comparing candidates consistently, quantifying downstream implications, and trusting rankings for discovery guidance remain difficult.

Industrial post-combustion CO\textsubscript{2} capture exemplifies these challenges. Candidate solvents must perform under absorption--desorption cycling with high CO\textsubscript{2} uptake, manageable viscosity (for mass transfer and operability), and favorable regeneration profiles. Consequently, solvent down-selection is inherently multi-objective, with the most useful screening signals being derived quantities (e.g., working capacity, enthalpy proxy) from predicted temperature-dependent solubility curves rather than single-point properties.

Recent machine learning approaches have improved IL property prediction, but two gaps remain critical. First, standard predictors degrade under scarce, noisy, uneven data; while transfer learning and message-passing architectures show promise in data-rich settings, descriptor--fingerprint MLPs with ensemble uncertainty quantification better handle irregularly sampled, multi-condition IL datasets—especially for probabilistic comparisons across derived quantities. Second, candidate ranking relies on ad hoc weighting or single metrics despite fundamental process trade-offs. We address these by treating data processing, predictive modeling, and physics-based post-processing as a coupled system that exposes uncertainty to directly inform multi-objective down-selection.

We present an AI-driven workflow integrating (i)~supervised property prediction, (ii)~physics-based thermodynamic post- processing, (iii)~uncertainty-aware multi-objective selection, and (iv)~synthesis feasibility filtering. We train descriptor--fingerprint neural predictors for viscosity and CO\textsubscript{2} solubility, quantify epistemic uncertainty via resampled ensembles, and derive working capacity and Van't~Hoff enthalpy proxies from predicted solubility curves. We perform probabilistic Pareto selection by evaluating dominance across ensemble predictive ``worlds,'' ranking ILs by Pareto-front frequency under uncertainty. Finally, automated retrosynthesis tools filter candidates lacking plausible synthesis routes, ensuring actionable candidates for experimental follow-up.

%% file: ICDM/sections/01_background.tex
\section{Related Work}

\begin{figure*}[t]
        \centering
        \includegraphics[width=0.85\linewidth]{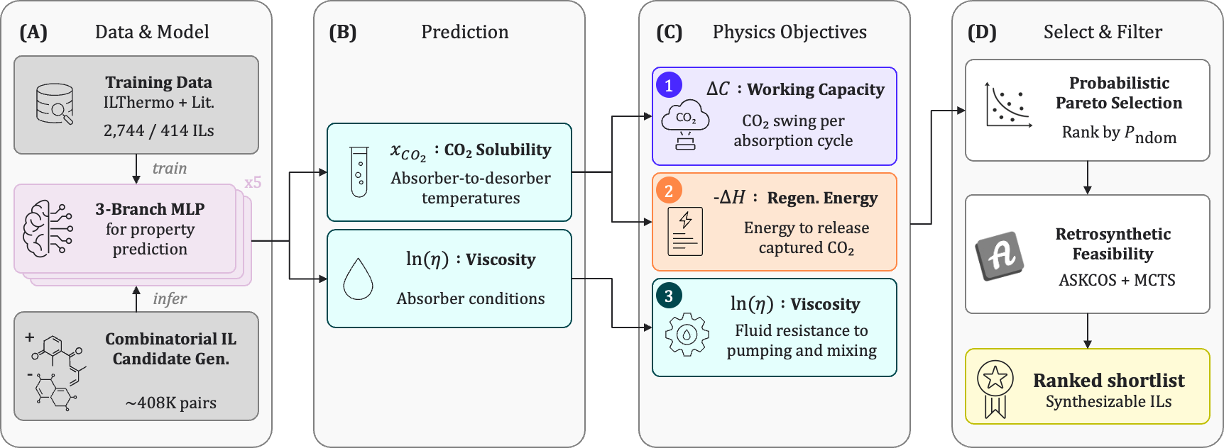}
        \caption{\textbf{Uncertainty-aware workflow for ionic liquid solvent discovery.}
        In \textbf{(A)}, the MLP is trained as a five-member resampled ensemble,
        each member defining one predictive ``world.'' In \textbf{(C)}, working
        capacity $\Delta C$ and the heat-of-absorption proxy $-\Delta H$ are derived
        from a Van't Hoff fit of the predicted temperature-dependent solubility
        curve. In \textbf{(D)}, candidates are filtered by objective-level
        feasibility cutoffs, then ranked by Pareto non-dominance probability
        $P_{\mathrm{ndom}}$ across the five worlds.}
        \label{fig:pipeline}
\end{figure*}

\paragraph{ML for ionic-liquid property prediction.}
Machine learning has been widely applied to predict IL properties from ion identity and operating conditions
\cite{Koutsoukos2021ILReview,Racki2025}. For CO$_2$ capture, supervised predictors of solubility have been developed using descriptor- and group-contribution-augmented models as well as neural architectures \cite{Song2020CO2IL,Ali2024CO2IL,Laakso2025SigmaProfileCO2}. Viscosity prediction combines structural features with temperature dependence via ANN-based and hybrid approaches \cite{Chen2022ViscosityILWater,Racki2025}. Lightweight, scalable predictors (e.g., descriptor--fingerprint MLPs) are effective for high-throughput screening when paired with careful preprocessing \cite{Koutsoukos2021ILReview}.

\paragraph{Uncertainty quantification via ensembles and resampling.}
Deep ensembles provide a competitive approach to epistemic uncertainty estimation \cite{Lakshminarayanan2017DeepEnsembles}. Our resampled-ensemble construction is motivated by classical resampling for variance estimation in finite-population sampling \cite{Gross1980,WangPengKim2022FinitePopBootstrap}. The delete-a-group / delete-a-fraction philosophy yields nearly unbiased estimates in stratified settings \cite{Kott2001DAGJK}, with without-replacement bootstrap being among the least biased for quantile variance estimation \cite{Saigo2010FourBootstraps}. This motivates using stable, repeated fits under systematic resampling as an uncertainty proxy during screening.

\paragraph{Multi-objective screening and Pareto methods under uncertainty.}
IL selection is inherently multi-objective, involving trade-offs between capacity, regeneration behavior, and viscosity. Pareto-based decision-making avoids arbitrary scalarization and is standard in materials design and active learning \cite{Jablonka2021BiasFreeMOAL}. Multi-objective Bayesian optimization explicitly accounts for uncertainty via acquisition functions integrating over noisy observations \cite{Daulton2021NEHVI}, with practical algorithms surveyed in \cite{Low2023MappingPareto}. Our probabilistic Pareto selection propagates predictive uncertainty into multi-objective decisions by evaluating dominance across ensemble-induced predictive ``worlds'', yielding robustness-oriented ranking.

\paragraph{Synthesis feasibility and retrosynthetic filtering.}
Top-ranked in silico candidates may be difficult to synthesize. While heuristic accessibility metrics provide fast approximations \cite{Ertl2009SAscore}, modern pipelines incorporate retrosynthetic planning using learned reaction models and search-based planners \cite{Jiang2023AIRetrosynthesisReview}. Recent work proposes retrosynthesis-guided scores incorporating route-level signals for better alignment with practical realizability \cite{Gao2025RetroScore}. We use retrosynthetic filtering as a final feasibility screen to ensure candidates remain actionable for experimental follow-up.

%% file: ICDM/sections/02_methods.tex
\section{Methods}
\label{sec:methods}

\subsection{Framework Overview}
\label{sec:framework}

We model temperature- and pressure-dependent CO$_2$ solubility and viscosity of ionic liquids as functions of molecular structure, trained on curated experimental datasets and deployed to prioritize IL candidates for post-combustion CO$_2$ capture in refinery operations.

Because experimental IL data is sparse and unevenly distributed across temperature--pressure space \cite{Mehrkesh2016, D3DD00040K, data4020088, SONG2020115752, Dong2007}, we learn structure--property relationships that generalize to untested conditions and novel chemistries, particularly for refinery scenarios (e.g., 1--2 bar, 40--120$^{\circ}$C) that may not align with literature measurements.

Our modular pipeline comprises five stages:
\begin{enumerate}
    \item \textbf{Ionic Liquid Candidate Space.} Over 400,000 IL candidates generated by combinatorial pairing of curated cations and anions.
    \item \textbf{MLP-based Property Prediction.} Descriptor--fingerprint neural networks predict CO$_2$ solubility and viscosity from cation/anion SMILES and thermodynamic conditions.
    \item \textbf{Physics-Based Thermodynamic Estimation.} Van't Hoff analysis derives working capacity and regeneration energy from predicted solubility curves.
    \item \textbf{Probabilistic Pareto Selection.} Candidates ranked by Pareto-front frequency across ensemble predictive ``worlds,'' optimizing working capacity $\Delta C$, heat of absorption $\Delta H$, and viscosity $\ln \eta$.
    \item \textbf{Synthesis Feasibility Filtering.} Automated retrosynthesis tools eliminate candidates lacking plausible synthesis routes.
\end{enumerate}

This framework enables in silico screening of hundreds of thousands of candidates without high-throughput experiments. Combining learned representations with physics-motivated scoring yields robust, interpretable decisions; hybrid ML--thermodynamic models outperform purely empirical or physics-based baselines \cite{D3DD00040K, Vermeire_2021}. Unlike prior work focusing on out-of-industrial conditions or handcrafted rules \cite{Racki2025}, our approach provides a data-driven, extensible solution for real-world refinery constraints.

\subsection{Ionic Liquid Candidate Space}
\label{sec:il-gen}

We generated ~408{,}000 ionic liquid (IL) candidates by combinatorially pairing cations and anions from cleaned datasets outlined in Section ~\ref{sec:mlp}. While this naive Cartesian approach constrains the search space to known ions, the framework is fully modular and can incorporate any generative strategy, including rule-based design, structural editing, or generative models to expand chemical diversity.

\subsection{MLP-based molecular property prediction}
\label{sec:mlp}

We model viscosity and CO$_2$ solubility as supervised regression from cation/anion SMILES and thermodynamic conditions. Viscosity predicts $\ln(\eta)$ from $T$; solubility predicts $x_{\mathrm{CO_2}} \in [0,1]$ from $T$ and $P$.

\paragraph{Datasets.}
We curate experimental viscosity and CO$_2$ solubility datasets from ILThermo and peer-reviewed literature; sources, coverage statistics, and filtering criteria are summarized in Table~\ref{tab:dataset-summary}, with full curation details in Appendix~\ref{sec:appendix-data}. Coverage is uneven across chemical and thermodynamic space: anion diversity is substantially sparser than cation diversity (e.g., only 107 anions vs.\ 159 cations across the 414 solubility ILs), and measurements are concentrated in well-studied imidazolium chemistries and near-ambient conditions, leaving much of the candidate space extrapolative.

\begin{table}[t]
\centering
\caption{\textbf{Dataset summary} for the viscosity and CO$_2$ solubility prediction targets, after curation. Filtering criteria are applied during curation (Appendix~\ref{sec:appendix-data}).}
\label{tab:dataset-summary}
\small
\setlength{\tabcolsep}{4pt}
\begin{tabular}{lll}
\toprule
 & \textbf{Viscosity $\ln(\eta)$} & \textbf{CO$_2$ solubility $x_{\mathrm{CO_2}}$} \\
\midrule
Sources & \multicolumn{2}{l}{ILThermo \cite{Dong2007} + literature \cite{VENKATRAMAN2017162, CHEN2025335, SONG2020115752}} \\
$N$ (measurements) & 20{,}309 & 17{,}261 \\
Unique ILs & 2{,}744 & 414 \\
Cations / anions & 1{,}218 / 315 & 159 / 107 \\
$T$ range (K) & $[253, 420]$ & $[243, 420]$ \\
$P$ range (bar) & --- (ambient) & $[0, 499.9]$ \\
\midrule
\multicolumn{3}{l}{\textbf{Filtering criteria}} \\
\multicolumn{3}{p{0.92\columnwidth}}{(i) Thermodynamic envelope: exclude $T > 420$ K, $P > 200$ bar. (ii) Duplicate resolution: average values within 5\% relative deviation; discard conflicting pairs. (iii) Physical consistency: enforce monotonic $x_{\mathrm{CO_2}}$ in $T$/$P$ and decreasing $\eta$ with $T$.} \\
\bottomrule
\end{tabular}
\end{table}

\paragraph{Molecular representations.}
We compute RDKit descriptors (232 features per ion; see Appendix~\ref{app:rdkit_features_used}) and Morgan fingerprints (ECFP4, 512 bits per ion), concatenating to form 464-dimensional descriptor and 1024-dimensional fingerprint vectors per IL pair. Physics-inspired condition encodings are $c_\eta(T) = [1000/T, T]$ for viscosity and $c_x(T,P) = [1000/T, T, P, \log(1+P)]$ for solubility.

\paragraph{Model architecture and training.}
A three-branch MLP processes descriptor, fingerprint, and condition features in separate subnetworks, then concatenates hidden states $h_d$, $h_f$, $h_c$ into a joint head. Branch architectures and training details (Huber loss, early stopping, pair-scaffold 5-fold splitting) are in Appendix~\ref{app:mlp_architecture}.

\begin{figure}[t]
        \centering
        \includegraphics[width=\linewidth]{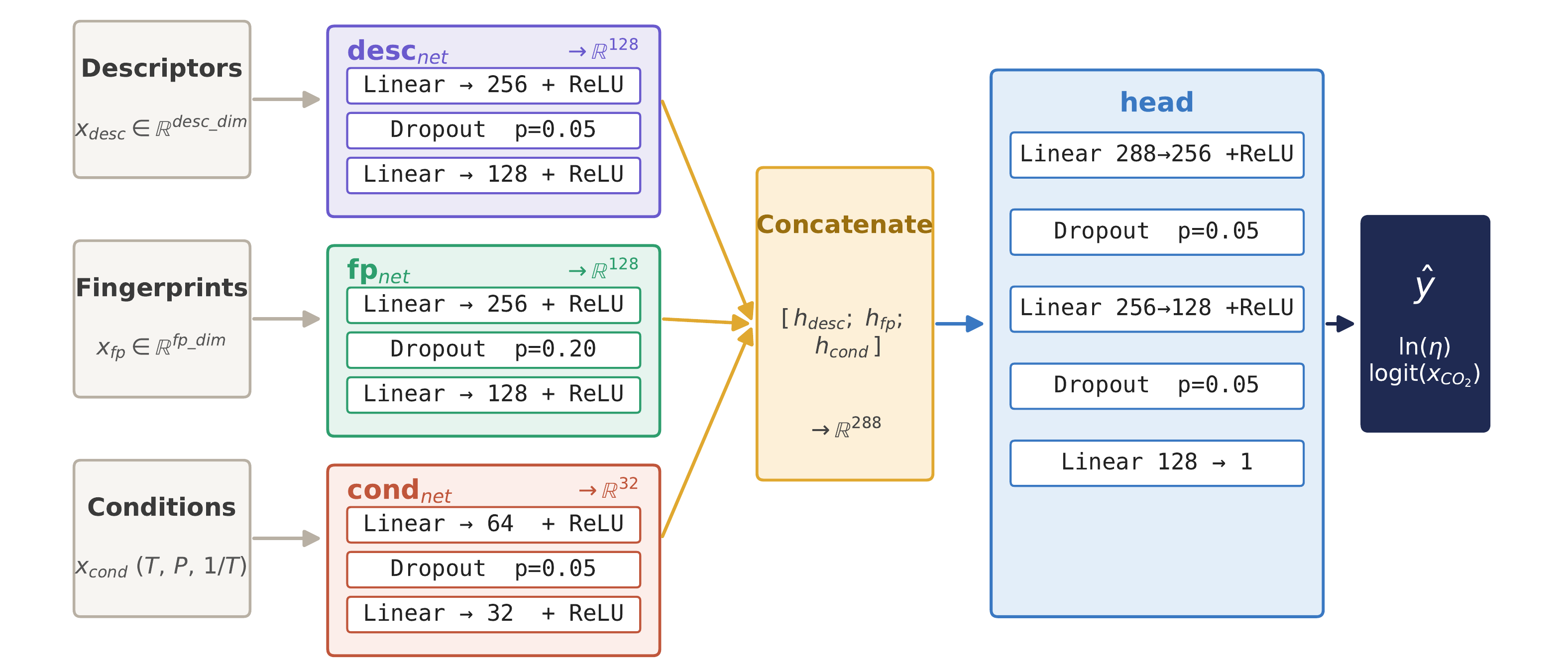}
        \caption{\textbf{Three-branch MLP architecture} for IL property prediction. Descriptor, Morgan-fingerprint, and thermodynamic-condition inputs are each processed by a separate subnetwork, then concatenated and passed through a shared head to predict the targets $\hat{y}$: $\ln(\eta)$ for viscosity or $\mathrm{logit}(x_{\mathrm{CO_2}})$ for solubility.}
\label{fig:mlp-architecture}
        \label{fig:model}
\end{figure}

\paragraph{Uncertainty quantification.}
We train ensembles of $B=5$ MLPs with identical architecture but different random seeds on subsampled training sets (90\% training / 10\% validation per member). We report ensemble mean as the point estimate and ensemble standard deviation as epistemic uncertainty proxy. This subsampled approach was selected over bootstrap alternatives based on sensitivity analyses showing stable rankings and lower computational cost. Full methodology is in Appendix~\ref{app:uncertainty_quantification_analysis}.

\subsection{Physics-Based Thermodynamic Property Estimation}
\label{sec:thermo}

IL performance for post-combustion CO$_2$ capture depends on absorption capacity and regeneration energy. We post-process predicted CO$_2$ solubilities using standard physical approximations to enable ranking and feasibility filtering.

For each candidate IL, we predict temperature-dependent CO$_2$ solubility at 20 evenly spaced points over $T \in [313.15, 393.15]$ K at 1 bar (representative of refinery stack pressure).

\paragraph{Working capacity.}
Practical solvent performance cannot be assessed from equilibrium absorption alone \cite{Raganati2020}. We define \textit{working capacity} as the difference between absorption and regeneration conditions:
\begin{equation}
\Delta C = C(T_\text{abs}) - C(T_\text{des}),
\end{equation}
where $T_\text{abs} = 313.15$ K (40$^{\circ}$C) and $T_\text{des} = 393.15$ K (120$^{\circ}$C) reflect realistic tower operating conditions at a typical refinery carbon capture unit. This excludes reversible physisorption at elevated temperatures.

\paragraph{Heat of absorption.}
The heat of absorption $\Delta H_{\text{abs,CO}_2}$ represents the enthalpic contribution to solvent regeneration and correlates strongly with overall regeneration energy. While total regeneration energy also includes sensible heat and steam generation effects \cite{Vinjarapu2024}, the heat of absorption provides a useful first-order proxy for screening candidate solvents. Because CO$_2$ absorption in ionic liquids is typically exothermic, $\Delta H_{\text{abs,CO}_2}$ values are negative. For clarity and ease of comparison, we report and visualize the negated quantity, $-\Delta H_{\text{abs,CO}_2}$, such that larger values correspond to stronger binding and higher regeneration energy, while smaller values indicate more favorable (lower-energy) regeneration behavior. All values are reported in kJ/mol unless stated otherwise. For reference, conventional amine solvents such as MEA require approximately 80--100~kJ/mol CO$_2$ for regeneration \cite{Rochelle2009, Vega2016}.

We estimate $\Delta H_{\text{abs,CO}_2}$ by fitting predicted temperature-dependent solubility data to a Van’t Hoff relation. The coefficient of determination $R^2$ is used to quantify fit quality. 


\subsection{Probabilistic Pareto Selection Under Ensemble Uncertainty}
\label{sec:pareto}

We rank IL candidates under predictive uncertainty by evaluating Pareto dominance across an ensemble of $B$ joint predictive ``worlds.'' We define a \emph{predictive world} $b \in \{1, \ldots, B\}$ as the joint set of predictions produced by ensemble member $b$: a single trained model yields, for every candidate $i$ and every objective, one joint realization $y_i^{(b)}$. Because all objectives for a given candidate are read from the same model, each world preserves the cross-objective correlations induced by that model rather than treating objectives as independent draws. We optimize three process-relevant objectives:
\begin{equation}
y_i^{(b)} = (\Delta C_i^{(b)}, \Delta H_i^{(b)}, \ln\eta_i^{(b)}),
\end{equation}
where $\Delta C$ is working capacity, $\Delta H$ is heat of absorption (Van't Hoff proxy), and $\ln\eta$ is log-viscosity at absorption temperature.



Candidate $i$ strictly dominates $j$ in world $b$ if it is superior or equal on all objectives with strict improvement on at least one. We define the non-dominance indicator:
\begin{equation}
f_i^{(b)} = \mathbb{I}[\nexists\, j \neq i \text{ s.t. } j \succ i \text{ in world } b],
\end{equation}
and score each candidate by its Pareto-front frequency across worlds:
\begin{equation}
\label{eq:pps}
P_{\mathrm{ndom}}(i) = \frac{1}{B}\sum_{b=1}^{B} f_i^{(b)} \in [0,1].
\end{equation}
Candidates are ranked by $P_{\mathrm{ndom}}(i)$. This approach propagates predictive uncertainty into multi-objective selection, enabling interpretable ranking without arbitrary scalarization.

\subsection{Synthesis Feasibility Screening}
\label{methods:synthesis-feas}

The final filter assesses the viability of laboratory and industrial synthesis pathways for top Pareto candidates. We employ retrosynthetic analysis using the ASKCOS platform enhanced with Monte Carlo Tree Search (MCTS) to efficiently explore synthetic routes. Synthetic feasibility is quantified by (i) average plausibility across all steps in the retrosynthetic tree, (ii) average tree depth, and (iii) maximum plausibility across all routes per IL. This ensures the final candidate shortlist reflects both chemical promise and manufacturability.

%% file: ICDM/sections/03_results.tex
\section{Results and Discussion}
\label{sec:results}

We present the results of our AI-guided pipeline for identifying and ranking ionic liquid (IL) candidates under typical refinery flue gas CO$_2$ capture conditions. We focus our discussion on two complementary results: (i) model predictive performance and (ii) Pareto top-performing candidate analysis.


\subsection{Model Performance and Predictive Confidence}
\label{sec:results-model}

We begin by evaluating the robustness and generalizability of our MLP-based models for predicting IL properties under a 5-fold balanced pair-scaffold cross-validation protocol and a pair-scaffold-disjoint held-out test split (Table~\ref{tab:baseline_perf}). We group by \emph{pair}-scaffold--the concatenated Bemis--Murcko scaffolds of the cation and anion (see Appendix~\ref{app:mlp_architecture}). For viscosity in natural-log units, $\ln(\eta)$, the MLP achieves a pooled out-of-fold CV MAE of $0.51 \pm 0.10$ with CV $R^2 = 0.77 \pm 0.14$, and a held-out test MAE of $0.50$ with test $R^2 = 0.86$. For CO$_2$ solubility as mole fraction, $x_{\mathrm{CO_2}}$, the MLP attains a CV MAE of $0.06 \pm 0.01$ with CV $R^2 = 0.83 \pm 0.05$, and a held-out test MAE of $0.03$ with test $R^2 = 0.98$. All values are reported as mean $\pm$ standard deviation over the five folds/trained fold models.

\input{tables/pred_benchmarks}

\paragraph{Baseline comparison} Under the same balanced pair-scaffold split protocol, we evaluated XGBoost, random forest, a Chemprop-based directed message passing neural network \cite{Heid2023}, ChemBERTa fine-tuning, and ILTransR fine-tuning baselines. Table~\ref{tab:baseline_perf} summarizes CV (pooled out-of-fold) and pair-scaffold-disjoint test performance across both targets.

\begin{figure}[t]
    \centering
    \includegraphics[width=\linewidth]{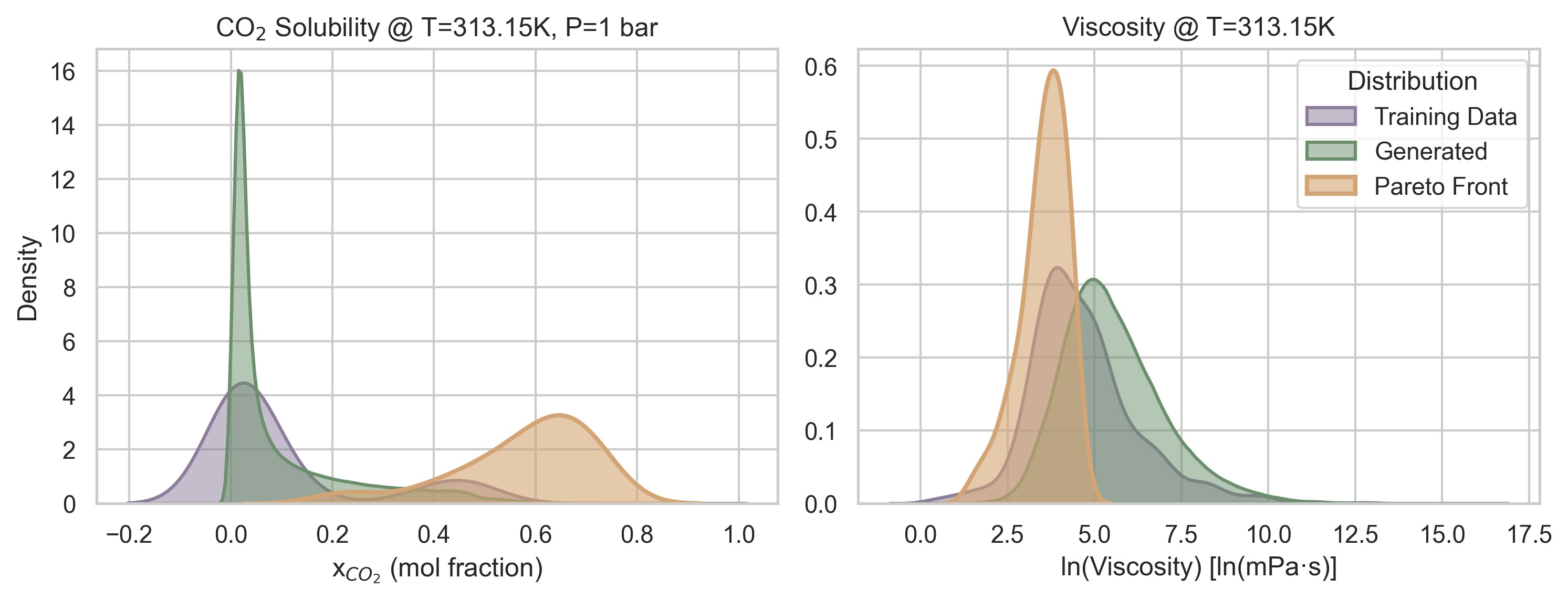}
    \caption{\textbf{Distributions of predicted properties.} CO$_2$ solubility and natural log viscosity properties for generated ionic liquids compared to those from raw training dataset. Plots show property distributions at absorption conditions ($T=313.15 \text{K}, P=1 \text{bar}$). Pareto front distribution represent isolated top performers and a subset of generated.}
    \label{fig:kpe-dists}
\end{figure}

\paragraph{Distribution comparison} We compare the property distributions with kernel density estimation plots in Figure \ref{fig:kpe-dists} to determine how well the MLP models the training data. The predicted property distributions of the generated ILs show moderate overlap with those from the training datasets for CO$_2$ solubility and viscosity indicating good predictive performance. Additionally, the distributions of the Pareto front candidates exhibit the shift towards desired properties: higher CO$_2$ solubility and lower viscosity respectively.

\subsection{Pareto Top-Performing Ionic Liquids Analysis}

\begin{figure*}[t]
        \centering
        \includegraphics[width=0.65\linewidth]{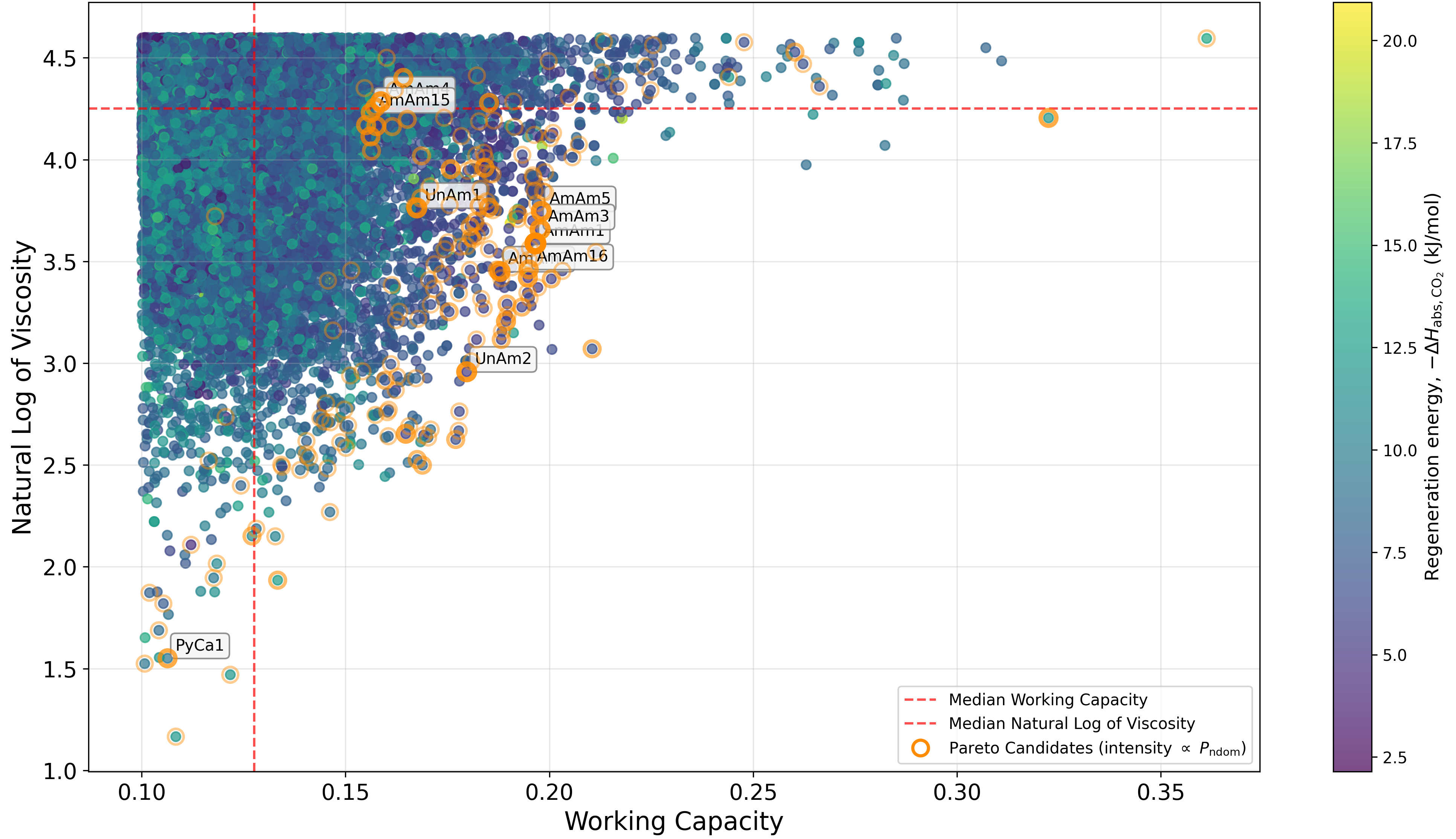}
        \caption{\textbf{Probabilistic Pareto front.}
        Working capacity versus ln-viscosity for filtered IL candidates ($N \approx 26{,}000$).
        Color indicates regeneration energy (kJ/mol CO$_2$).
        Orange circles highlight candidates appearing on the probabilistic Pareto front.
        Circle intensity is proportional to the Pareto non-dominance probability, $P_{\mathrm{ndom}}$, reflecting the fraction of ensemble worlds in which a candidate is Pareto-optimal.
        Labeled candidates are top-ranked ILs (see Table~\ref{tab:pareto_candidates})
        }
        \label{fig:work-cap-visc}
\end{figure*}

A central challenge in solvent design for post-combustion CO$_2$ capture is managing the inherent trade-off between absorption strength and fluid operability: candidates with high CO$_2$ uptake often exhibit prohibitively high viscosity at operating conditions. We analyze how the proposed uncertainty-aware Pareto framework resolves this trade-off and surfaces robust, high-performing ionic liquid (IL) candidates.

We evaluated the $\sim$408{,}000 generated IL candidates across five ensemble predictive worlds. In each world, we apply objective-level feasibility filtering prior to Pareto selection, removing candidates whose predicted performance is infeasible under industrial operating conditions. Specifically, we exclude ILs with low working capacity ($\Delta C < 0.10$), high viscosity ($\ln(\eta) > 4.6$), or weak CO$_2$ binding as indicated by near-zero heat of absorption ($-\Delta H_{\mathrm{abs,CO_2}} < 5$~kJ/mol). These criteria remove candidates that fail to meaningfully compete with benchmark amine solvents, exceed practical viscosity limits for process equipment, or exhibit insufficient binding strength and absorption kinetics.

The probabilistic Pareto selection procedure described in Section~\ref{sec:pareto} is then applied to the remaining candidates. Candidates are ranked by their Pareto non-dominance probability, $P_{\mathrm{ndom}}$, which captures the consistency of Pareto-optimal performance under model uncertainty. We focus our analysis on high-confidence candidates with $P_{\mathrm{ndom}} \geq 0.4$.

\paragraph{Probabilistic Pareto Front.} Figure~\ref{fig:work-cap-visc} plots working capacity versus ln-viscosity for filtered candidates, with Pareto front members highlighted and color intensity indicating $P_{\mathrm{ndom}}$. The distribution reveals a clear tension: high-capacity candidates ($\Delta C > 0.20$) predominantly occupy higher-viscosity regimes while low-viscosity candidates sacrifice capacity, reflecting the physicochemical coupling between strong CO$_2$ binding and reduced fluid mobility.

Accounting for predictive uncertainty across ensemble worlds, candidates with high $P_{\mathrm{ndom}}$ form a compact subset of the front that consistently balances capacity and viscosity across model realizations, rather than excelling in a single objective under isolated predictions; candidates at the extremes exhibit lower $P_{\mathrm{ndom}}$. The most robust candidates cluster in the lower-right quadrant—above-median capacity with below-median viscosity—an attractive regime for practical deployment.

\begin{figure*}[t]
    \centering
    \subfloat[Working capacity, $\Delta C$ $\uparrow$]{
        \includegraphics[width=0.48\linewidth]{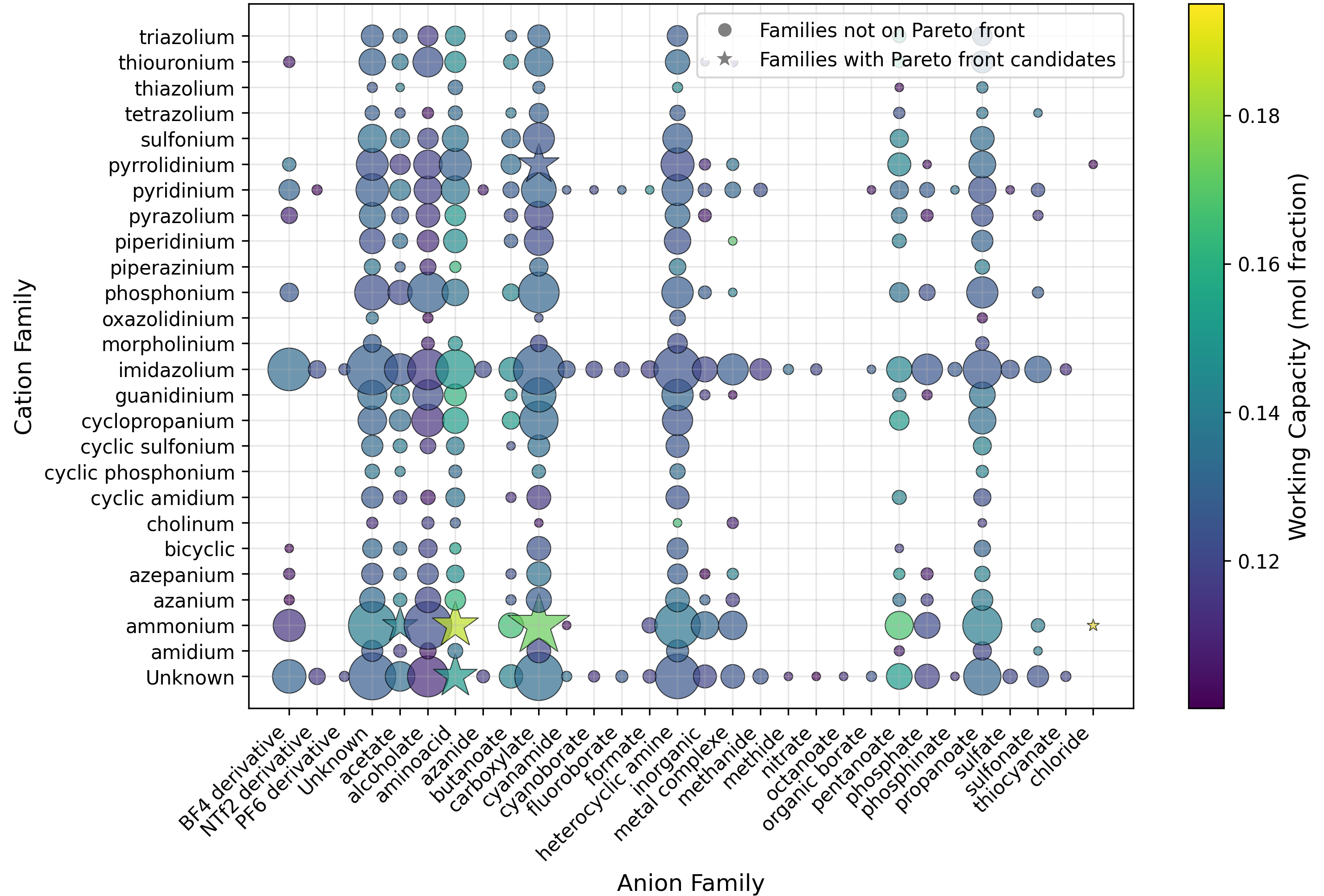}
        \label{fig:family-performance-cap}
    }
    \hfil
    \subfloat[Viscosity, $\ln(\eta)$ $\downarrow$]{
        \includegraphics[width=0.48\linewidth]{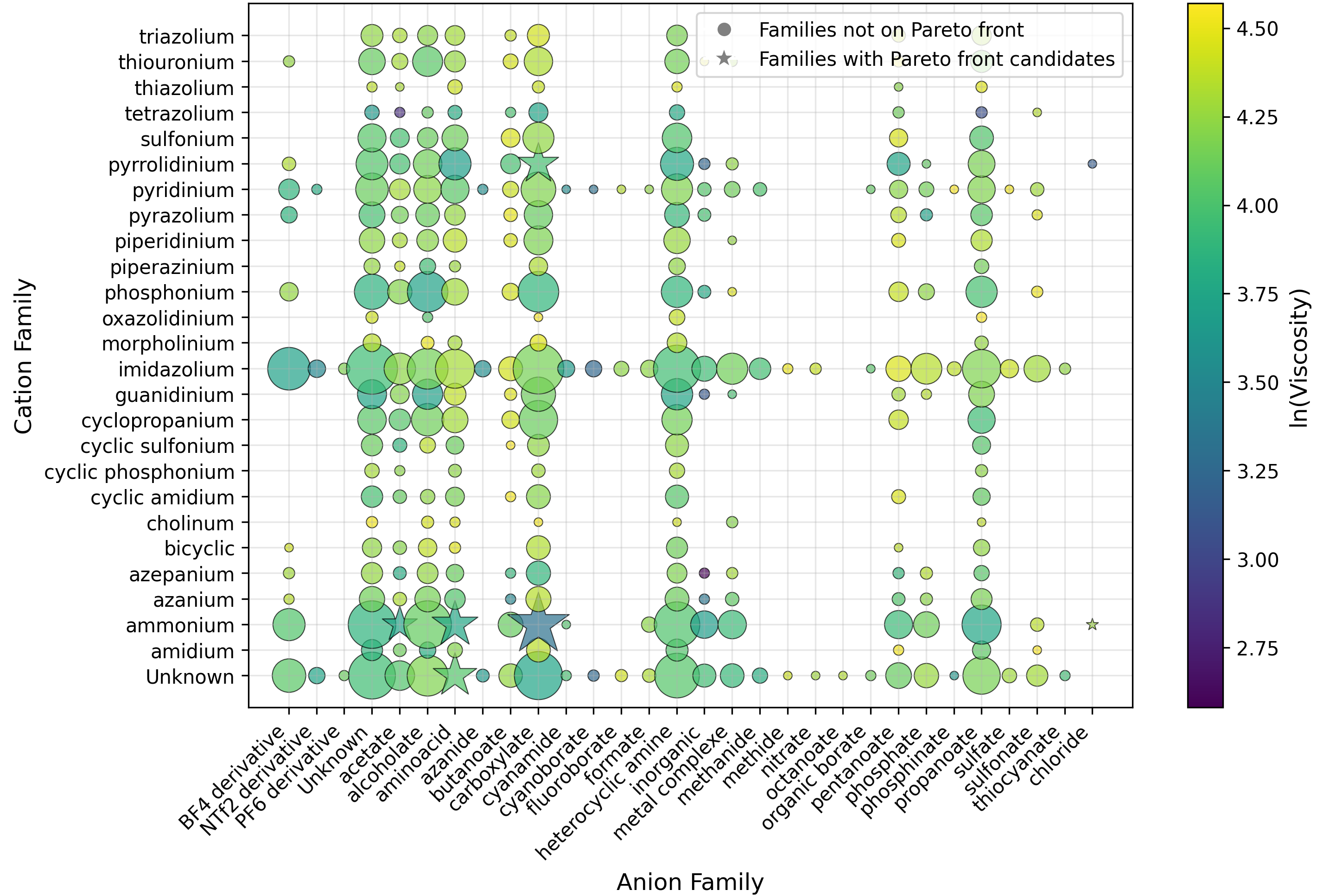}
        \label{fig:family-performance-visc}
    }
    \caption{\textbf{Family-level aggregation} of filtered ionic liquid candidates, grouped by cation and anion families. Stars indicate family combinations with at least one candidate appearing on the probabilistic Pareto front ($P_{\mathrm{ndom}} \geq 0.4$). Bubble size is proportional to the number of candidate ILs within each family combination. To avoid bias toward families with larger candidate pools, values are computed using only the top 50 performers per family (for respective metric). Arrow indicates whether higher or lower is better. (\textbf{a}) Average working capacity, highlighting families that achieve large absorption–desorption swings. (\textbf{b}) Average ln-viscosity, highlighting families that favor fluid operability under absorption conditions.}
    \label{fig:family-bubble-summary}
\end{figure*}

\paragraph{Family-Level Insights and Chemical Trends}
To contextualize these Pareto-optimal candidates chemically, we grouped ILs by cation and anion family and aggregated performance across families (Figure~\ref{fig:family-bubble-summary}). The working capacity plot (Figure~\ref{fig:family-performance-cap}) shows that amino acid anion-based ILs tend to achieve high capacity, reflecting their strong, often reversible interactions with CO$_2$. Among these, ammonium cations dominate the highest-performing combinations, with the ammonium–amino acid (AmAm) family accounting for 8 of the top 10 Pareto-ranked candidates.

In contrast, the viscosity aggregation (Figure~\ref{fig:family-performance-visc}) highlights a complementary trend: families that achieve the lowest viscosities are not always those with the highest capacities. Pyrrolidinium- and certain quaternary ammonium-based families tend to favor fluid operability, whereas imidazolium-based ILs, while extensively studied in the literature, do not uniquely dominate the most favorable regions when both objectives are considered jointly.

Taken together, these plots illustrate that Pareto-optimal performance arises not from families that maximize a single metric, but from those that achieve a balanced compromise between capacity and viscosity. The presence of stars across multiple family combinations further indicates that robust Pareto candidates are distributed across diverse chemistries rather than concentrated in a single dominant family.

Several high-confidence Pareto candidates fall outside predefined family templates and are labeled as ``Unknown'' (UnAm in Figure~\ref{fig:work-cap-visc}). These candidates correspond to rare or unconventional cation motifs that nonetheless achieve stable multi-objective performance. Their consistent appearance across ensemble worlds highlights the ability of uncertainty-aware Pareto selection to surface promising but underexplored chemistries that would likely be overlooked by heuristic or single-metric screening approaches.

\paragraph{Survey of Top-Performing Candidate ILs}

\input{tables/top-candidates}

Table~\ref{tab:pareto_candidates} reports the top 10 candidates ranked by $P_{\mathrm{ndom}}$. The highest-ranked candidate (AmAm1) appears on the Pareto front in all five ensemble worlds ($P_{\mathrm{ndom}} = 1.0$). Across the top 10, ammonium–amino acid ILs dominate, exhibiting working capacities of 0.16–0.20 and ln-viscosities corresponding to $\sim$20–70 mPa${\cdot}$s. Of the top-performing candidate ILs (Pareto candidates), retrosynthetic analysis shows that 6\% possess synthesis routes with average plausibility scores across routes of over 90\%.

Candidate PyCa1 (pyrrolidinium–carboxylate) illustrates the opposing extreme, achieving the lowest viscosity among top candidates (ln$\eta \approx 1.55$, $\sim$5 mPa${\cdot}$s) at the cost of reduced working capacity. The presence of structurally uncharacterized UnAm candidates among the top ranks further emphasizes the ability of uncertainty-aware Pareto selection to identify robust yet unconventional solvent chemistries.

Reported standard deviations reflect cross-model uncertainty. For most top candidates, $\Delta C$ varies by $\sim \pm 0.05$ and ln-viscosity by $\sim \pm 0.3$ across ensemble worlds, indicating moderate predictive variance. Candidates with smaller dispersion (e.g., PyCa1) exhibit more stable rankings, suggesting lower model sensitivity.

%% file: tables/pred_benchmarks.tex
\begin{table}[t]
\setlength{\tabcolsep}{3pt}
\centering
\caption{\textbf{Baseline predictive performance.} Baseline performance under 5-fold balanced pair-scaffold cross-validation (CV) and pair-scaffold-disjoint held-out test evaluation for both targets: $\ln(\eta)$ and $x_{\mathrm{CO_2}}$. CV metrics are computed from pooled out-of-fold predictions across the five folds. CV metrics are reported as mean $\pm$ standard deviation across the five folds; held-out test metrics are reported as single evaluations on the pair-scaffold-disjoint test set.}
\label{tab:baseline_perf}
\begin{minipage}{\columnwidth}
\centering
\footnotesize
\textbf{Viscosity} $\ln(\eta)$
\vspace{3pt}
\begin{tabular}{l r r r r}
\toprule
Model & CV MAE & CV $R^2$ & Test MAE & Test $R^2$ \\
\midrule
XGBoost & $0.48_{\pm 0.08}$ & $0.74_{\pm 0.10}$ & $0.51$ & 0.81 \\
Random Forest & $0.61_{\pm 0.03}$ & $0.86_{\pm 0.06}$ & $0.63$ & 0.74 \\
ExtraTrees & $0.65_{\pm 0.01}$ & $0.89_{\pm 0.03}$ & $0.71$ & 0.72 \\
Chemprop & $0.61_{\pm 0.12}$ & $0.87_{\pm 0.20}$ & $0.53$ & 0.69 \\
ChemBERTa (finetune) & $0.64_{\pm 0.07}$ & $0.91_{\pm 0.09}$ & $0.70$ & 0.71 \\
ILTransR (finetune) & $0.46_{\pm 0.07}$ & $0.84_{\pm 0.06}$ & $0.51$ & $0.78$ \\
\textbf{MLP (final method)} & $\mathbf{0.51_{\pm 0.10}}$ & $\mathbf{0.77_{\pm 0.14}}$ & $\mathbf{0.50}$ & \textbf{0.86} \\
\bottomrule
\end{tabular}
\vspace{8pt}

\textbf{CO$_2$ solubility} $x_{\mathrm{CO_2}}$
\vspace{3pt}
\begin{tabular}{l r r r r}
\toprule
Model & CV MAE & CV $R^2$ & Test MAE & Test $R^2$ \\
\midrule
XGBoost & $0.04_{\pm 0.01}$ & $0.89_{\pm 0.05}$ & $0.07$ & 0.89 \\
Random Forest & $0.09_{\pm 0.02}$ & $0.62_{\pm 0.03}$ & $0.14$ & 0.63 \\
ExtraTrees & $0.09_{\pm 0.03}$ & $0.62_{\pm 0.11}$ & $0.14$ & 0.62 \\
Chemprop & $0.18_{\pm 0.04}$ & $0.88_{\pm 0.02}$ & $0.27$ & 0.85 \\
ChemBERTa (finetune) & $0.14_{\pm 0.04}$ & $0.28_{\pm 0.25}$ & $0.19$ & 0.29 \\
ILTransR (finetune) & $0.05_{\pm 0.01}$ & $0.86_{\pm 0.05}$ & $0.08$ & $0.90$ \\
\textbf{MLP (final method)}& $\mathbf{0.06_{\pm 0.01}}$& $\mathbf{0.83_{\pm 0.05}}$ & $\mathbf{0.03}$& \textbf{0.98} \\
\bottomrule
\end{tabular}
\end{minipage}
\end{table}

%% file: tables/top-candidates.tex
\begin{table*}[t]
\centering
\caption{\textbf{Selected Pareto front candidates.} Top section shows top 10 anonymized high-confidence candidates ($P_{\mathrm{ndom}} \geq 0.6$); bottom section reveals structures for ranks 91–100 ($P_{\mathrm{ndom}} = 0.2$). All metrics reported as mean $\pm$ standard deviation across 5 ensemble worlds.}
\label{tab:pareto_candidates}
\scriptsize
\begin{tabular}{clccccc}
\toprule
Rank & IL ID/Name & $P_{\mathrm{ndom}}$ & $\Delta C$ $\uparrow$ & $x_{\mathrm{CO}_2}^{\mathrm{abs}}$\textsuperscript{*} $\uparrow$ & ln($\eta$) $\downarrow$ & -$\Delta H_{\text{abs,CO}_2}$ $\downarrow$ \\
\midrule
\midrule
1  & AmAm1 & 1.0 & 0.197 $\pm$ 0.068 & 0.607 $\pm$ 0.050 & 3.59 $\pm$ 0.32 & $5.23$ $\pm$ 2.03 \\
2  & AmAm2 & 0.8 & 0.198 $\pm$ 0.055 & 0.629 $\pm$ 0.042 & 3.75 $\pm$ 0.22 & $5.01$ $\pm$ 1.64 \\
3  & AmAm3 & 0.8 & 0.188 $\pm$ 0.059 & 0.613 $\pm$ 0.048 & 3.45 $\pm$ 0.33 & $4.78$ $\pm$ 1.49 \\
4  & UnAm1 & 0.8 & 0.168 $\pm$ 0.063 & 0.653 $\pm$ 0.049 & 3.76 $\pm$ 0.18 & $3.85$ $\pm$ 1.61 \\
5  & AmAm4 & 0.8 & 0.198 $\pm$ 0.061 & 0.622 $\pm$ 0.045 & 3.66 $\pm$ 0.29 & $5.08$ $\pm$ 1.80 \\
6  & UnAm2 & 0.8 & 0.180 $\pm$ 0.062 & 0.545 $\pm$ 0.064 & 2.96 $\pm$ 0.27 & $5.22$ $\pm$ 1.70 \\
7  & AmAm5 & 0.8 & 0.159 $\pm$ 0.057 & 0.729 $\pm$ 0.034 & 4.28 $\pm$ 0.21 & $3.23$ $\pm$ 1.44 \\
8  & AmAm6 & 0.6 & 0.195 $\pm$ 0.059 & 0.576 $\pm$ 0.046 & 3.46 $\pm$ 0.36 & $5.47$ $\pm$ 1.80 \\
9  & AmAm7 & 0.6 & 0.156 $\pm$ 0.057 & 0.726 $\pm$ 0.033 & 4.23 $\pm$ 0.21 & $3.21$ $\pm$ 1.46 \\
10 & PyCa1 & 0.6 & 0.106 $\pm$ 0.005 & 0.213 $\pm$ 0.019 & 1.55 $\pm$ 0.17 & $9.36$ $\pm$ 1.94 \\
\midrule
\midrule
91 & Octadecyl-dimethyl-cyclohexyl ammonium / $\beta$-Alaninate & 0.2 & 0.201 $\pm$ 0.042 & 0.631 $\pm$ 0.043 & 4.13 $\pm$ 0.24 & $5.04$ $\pm$ 1.41  \\
92 & Decyl-dimethyl-cyclohexyl ammonium / Butyrate & 0.2 & 0.180 $\pm$ 0.053 & 0.461 $\pm$ 0.046 & 3.57 $\pm$ 0.26 & $6.52$ $\pm$ 1.87  \\
93 & Triethyl-methoxymethyl ammonium / Hydroxide & 0.2 & 0.102 $\pm$ 0.002 & 0.224 $\pm$ 0.012 & 1.87 $\pm$ 0.11 & $7.91$ $\pm$ 0.07  \\
94 & Octadecyl-dimethyl-cyclohexyl ammonium / Glycinate & 0.2 & 0.196 $\pm$ 0.038 & 0.580 $\pm$ 0.038 & 4.12 $\pm$ 0.21 & $5.45$ $\pm$ 1.33  \\
95 & Ethyl-dimethyl-cyclohexyl ammonium / Leucinate & 0.2 & 0.185 $\pm$ 0.051 & 0.514 $\pm$ 0.033 & 3.79 $\pm$ 0.29 & $5.89$ $\pm$ 1.74  \\
96 & 1-Methoxymethyl-3-methylimidazolium / Tricyanomethanide & 0.2 & 0.121 $\pm$ 0.016 & 0.441 $\pm$ 0.032 & 2.74 $\pm$ 0.21 & $4.30$ $\pm$ 0.83 \\
97 & Trimethyl-methoxymethyl ammonium / Hydroxide & 0.2 & 0.105 & 0.244 $\pm$ 0.014 & 1.82 $\pm$ 0.09 & $7.48$ \\
98 & Octyl-dimethyl-cyclohexyl ammonium / Valinate & 0.2 & 0.207 $\pm$ 0.071 & 0.508 $\pm$ 0.030 & 4.07 $\pm$ 0.33 & $7.00$ $\pm$ 2.46 \\
99 & Heptyl-dimethylamino-cyclohexyl ammonium / Norvalinate & 0.2 & 0.196 $\pm$ 0.063 & 0.497 $\pm$ 0.030 & 3.52 $\pm$ 0.28 & $6.58$ $\pm$ 2.08 \\
100 & 1-Cyclopentyl-3-methylimidazolium / Glycinate & 0.2 & 0.181 $\pm$ 0.040 & 0.485 $\pm$ 0.030 & 3.46 $\pm$ 0.25 & $6.08$ $\pm$ 1.53 \\
\bottomrule
\end{tabular}
\\[2pt]
{\footnotesize \textsuperscript{*}$x_{\mathrm{CO}_2}^{\mathrm{abs}}$ is the predicted CO$_2$ mole fraction at absorption conditions ($T = 313.15$ K, $P = 1$ bar).}
\end{table*}

%% file: ICDM/sections/99_conclusions.tex
\section{Conclusion}

This work presents an end-to-end, modular framework for data-driven discovery of ionic liquid solvents for CO$_2$ capture under industrial operating conditions. Our approach integrates supervised property prediction, physics-based thermodynamic post-processing, uncertainty-aware multi-objective selection, and synthesis feasibility filtering into a coupled pipeline that prioritizes actionable candidates.

We screened over 400,000 candidate ILs and identified 46 high-performing compounds on the probabilistic Pareto front (P$_{\mathrm{ndom}} \geq 0.4$), of which 6\% possess synthesis routes with average plausibility scores across routes of over 90\%. These candidates achieve working capacities of 0.11--0.32 mol/mol, regeneration enthalpies ranging from $-3.2$ to $-12$ kJ/mol CO$_2$, and viscosities below 93 mPa$\cdot$s—properties that compare favorably with conventional solvents like MEA and MDEA. The best performers consistently pair amino acid or carboxylate anions with ammonium cations, validating known structure--property trends while identifying novel, unexplored combinations. Notably, 8 of the top 10 candidates belong to the ammonium--amino acid family, with working capacities exceeding 0.16 and regeneration energies of 3--5 kJ/mol CO$_2$, demonstrating the favorable balance of high capacity and moderate binding energy achievable with these pairings.

By coupling learned property predictions with physics-informed scoring and uncertainty quantification, our framework enables interpretable, multi-objective ranking without arbitrary scalarization. This supports flexible prioritization based on real-world process constraints and provides a transparent foundation for experimental validation. The modularity of our approach extends naturally to other properties, thermodynamic conditions, and discovery targets.

\emph{Future work.} Future work will expand the thermodynamic model to account for loading- and pressure-dependent effects, incorporate constraints on thermal stability and environmental toxicity, and close the discovery loop with experimental measurements on top-ranked candidates.

%% file: ICDM/sections/04_limitations.tex
\section{Limitations and Ethical Considerations} 
 Our predictions rely on data-driven models trained on sparse, noisy experimental records; while ensemble uncertainty quantification mitigates this limitation, experimental validation remains essential. Additionally, thermodynamic post-processing assumes equilibrium behavior and constant enthalpy across the evaluated temperature range, which may not hold under dynamic process conditions or for ILs with nonlinear solubility profiles.

%% file: ICDM/sections/genai.tex
\section*{GenAI Disclosure}
We used a generative AI tool to assist with writing by improving clarity and readability through
light editing (e.g., rephrasing and grammar/style suggestions). The tool was not used to generate
substantive new scientific content, results, data, figures, tables, or code. All AI-assisted edits
were reviewed and finalized by the authors, who take full responsibility for the content of this
paper.

%% file: ICDM/sections/999_appendix.tex
\section{Data Curation \& Processing Details}
\label{sec:appendix-data}

\paragraph{Data sources.}
Experimental CO$_2$ solubility and viscosity data for ionic liquids (ILs) were compiled from a combination of peer-reviewed literature sources and the NIST Ionic Liquids Database (ILThermo). Specifically, data were extracted from three comprehensive experimental studies covering a broad range of IL chemistries and thermodynamic conditions \cite{VENKATRAMAN2017162, CHEN2025335, SONG2020115752}, and supplemented with curated measurements from ILThermo \cite{Dong2007}.

All records were standardized to a common representation, including canonical SMILES for cations and anions, consistent temperature (K) and pressure (bar) units, and unified property definitions. When necessary, reported values were converted from mass- or molar-based units to mole fraction or natural-log viscosity to ensure consistency across datasets.

\paragraph{Duplicate and conflict resolution.}
Due to overlap across literature sources and ILThermo, duplicate measurements were identified based on identical $(\text{cation}, \text{anion}, T, P)$ tuples. When multiple values were reported for the same condition, we compared measurements pairwise:
\begin{itemize}
    \item If the relative deviation between reported values was less than 5\%, the measurements were averaged.
    \item If the relative deviation exceeded 5\%, the corresponding data point was discarded to avoid propagating inconsistent or potentially erroneous measurements.
\end{itemize}

\paragraph{Thermodynamic envelope filtering.}
To focus model training on conditions relevant to industrial post-combustion CO$_2$ capture, we excluded measurements outside a practical operating envelope:
\begin{itemize}
    \item Temperature $T > 420$ K
    \item Pressure $P > 200$ bar
\end{itemize}
These thresholds remove extreme laboratory conditions unlikely to be encountered in refinery or flue-gas capture settings, while retaining sufficient coverage for learning temperature- and pressure-dependent trends.

\paragraph{Physical consistency checks.}
We applied domain-informed consistency filters to remove physically implausible records:
\begin{itemize}
    \item \textbf{CO$_2$ solubility:} At fixed pressure, solubility should decrease monotonically with increasing temperature; at fixed temperature, solubility should increase monotonically with pressure. Measurements violating either trend were removed, as they contradict expected Henry’s law behavior in the relevant regime.
    \item \textbf{Viscosity:} Viscosity should decrease with increasing temperature for ionic liquids under single-phase conditions. Records exhibiting increasing viscosity with temperature were discarded, except in rare cases associated with reported phase transitions.
\end{itemize}

These filtering steps reduce noise from transcription errors, inconsistent experimental protocols, and outlier measurements, yielding a curated dataset that better reflects physically meaningful structure–property relationships.

\section{MLP-Based Property Prediction}

\subsection{RDKit features used}
\label{app:rdkit_features_used}

For each ion SMILES, we compute the full RDKit continuous descriptor list (\texttt{Descriptors.\_descList}; 217 descriptors per ion) and augment it with 15 chemically interpretable count features (Table~\ref{tab:rdkit_extra_counts}). This yields 232 descriptor features per ion. For an ionic liquid (IL) pair, we concatenate the cation and anion descriptor vectors to obtain a 464-dimensional descriptor vector. In implementation, pair-level descriptor feature names are formed by prefixing each per-ion feature with \texttt{cat\_} and \texttt{ani\_} (e.g., \texttt{cat\_MolWt}, \texttt{ani\_MolWt}).

\paragraph{RDKit descriptor families (217 features).}
The RDKit descriptor set includes: E-State indices; QED; molecular weight and electronic descriptors; BCUT2D descriptors; topological indices (BalabanJ, BertzCT, Ipc, Kier--Hall); surface area and polarity descriptors (LabuteASA, TPSA); VSA descriptors (PEOE\_VSA, SMR\_VSA, SlogP\_VSA, EState\_VSA, VSA\_EState); ring and heteroatom counts; and a library of functional-group fragment counts (\texttt{fr\_*}). The complete, ordered list is exactly RDKit's \texttt{Descriptors.\_descList} (217 entries); the 15 appended count features are given in Table~\ref{tab:rdkit_extra_counts}.

\begin{table}[t]
\centering
\caption{Additional per-ion count features appended to the RDKit descriptor vector (15 features).}
\label{tab:rdkit_extra_counts}
\begin{tabularx}{0.95\linewidth}{@{}lX@{}}
\toprule
\textbf{Category} & \textbf{Features} \\
\midrule
Element counts &
\texttt{num\_B, num\_C, num\_N, num\_O, num\_F, num\_Si, num\_P, num\_S, num\_Cl, num\_Br, num\_I, num\_Se} \\
Aggregates &
\texttt{num\_halogen} (F+Cl+Br+I) \\
Molecular counts &
\texttt{formal\_charge}, \texttt{heavy\_atom\_count} \\
\bottomrule
\end{tabularx}
\end{table}

\subsection{Feature preprocessing and target transformations}
\label{app:pred_data_preprocessing}

All preprocessing steps are fit on the training partition only and applied unchanged to validation/test/inference data to prevent leakage.

\paragraph{Viscosity target.}
We train directly on $y_\eta = \ln(\eta)$.

\paragraph{Solubility target.}
To improve numerical stability near boundaries, we train in logit space:
\begin{equation}
y_x = \mathrm{logit}(x_{\mathrm{CO_2}}) = \log\left(\frac{x_{\mathrm{CO_2}}}{1-x_{\mathrm{CO_2}}}\right),
\end{equation}
with clipping $x_{\mathrm{CO_2}} \leftarrow \mathrm{clip}(x_{\mathrm{CO_2}}, \varepsilon, 1-\varepsilon)$ and $\varepsilon = 10^{-6}$. At inference, we map back via sigmoid:
\begin{equation}
\hat{x}_{\mathrm{CO_2}} = \sigma(\hat{y}_x) = \frac{1}{1+\exp(-\hat{y}_x)}.
\end{equation}

\paragraph{Descriptor preprocessing.}
We remove degenerate (zero-variance) features, impute missing values with training-set median, and perform univariate feature selection on the descriptor matrix using an F-test for regression, retaining the top $k=300$ descriptor features. Selected descriptors and condition features are standardized via z-scoring using training-set means and standard deviations; Morgan fingerprint bits are not standardized.

\subsection{Neural network architecture and training}
\label{app:mlp_architecture}

\paragraph{Three-branch MLP.}
Let $d \in \mathbb{R}^k$ denote the selected descriptor vector ($k=300$), $f \in \mathbb{R}^{1024}$ the concatenated fingerprint, and $c$ the standardized condition vector (2D for viscosity, 4D for solubility). The model computes:
\begin{align}
h_d &= \mathrm{MLP}_d(d), \\
h_f &= \mathrm{MLP}_f(f), \\
h_c &= \mathrm{MLP}_c(c), \\
\hat{y} &= \mathrm{MLP}_{\mathrm{head}}([h_d \,\|\, h_f \,\|\, h_c]),
\end{align}
where $\|$ denotes concatenation. All hidden layers use ReLU activations.

Branch widths:
\begin{itemize}
    \item Descriptor branch: $k \to 256 \to 128$
    \item Fingerprint branch: $1024 \to 256 \to 128$
    \item Condition branch: $2 \to 32 \to 16$ (viscosity) or $4 \to 64 \to 16$ (solubility)
    \item Head: $(128+128+16) \to 128 \to 64 \to 1$
\end{itemize}
Dropout is applied as specified in the implementation.

\paragraph{Training.}
Models are trained with Smooth L1 (Huber) loss ($\delta=0.5$) using Adam optimization, mini-batching, and gradient clipping (max $\ell_2$-norm 5.0). Batch sizes: 2048 for viscosity, 4096 for solubility. Early stopping based on validation RMSE with property-specific patience and epoch caps.

\paragraph{Splitting protocol.}
We compute Bemis--Murcko scaffolds for each ion and define a pair-scaffold label by concatenating cation and anion scaffold SMILES. Balanced 5-fold splits are constructed over pair-scaffold groups, ensuring all datapoints sharing a pair scaffold remain in the same fold. This estimates generalization beyond near-duplicate chemistry.

\subsection{Uncertainty quantification and analysis}
\label{app:uncertainty_quantification_analysis}

\subsubsection{Training, ensembling, and uncertainty quantification methods}
\paragraph{Base predictor.}
All reported uncertainty analyses use the three-branch MLP described above (descriptor branch, fingerprint branch,
and condition branch, followed by a fully-connected head).
For viscosity, the model predicts $\ln(\eta)$ directly.
For CO$_2$ solubility, the model predicts $\mathrm{logit}(x_{\mathrm{CO_2}})$ and we report
$x_{\mathrm{CO_2}}=\sigma(\mathrm{logit}(x_{\mathrm{CO_2}}))$.

\paragraph{Resampled ensembles for epistemic uncertainty.}
To estimate \emph{epistemic} uncertainty (model uncertainty due to finite data), we trained an ensemble of $M=5$
independently initialized MLPs per task. Each ensemble member uses the same architecture and hyperparameters but is
trained on a different resampled training set. We evaluated two resampling schemes:
(1) \textbf{Subsampled ensemble (``dropout90/val10'').}
  For each member $m$, we sample $90\%$ of the available training records \emph{without replacement} to form the
  training subset and use the remaining $10\%$ as a holdout validation set for early stopping. This is conceptually
  related to delete-a-group / delete-a-fraction resampling, where replicated fits quantify estimator variability under
  systematic deletion~\cite{Kott2001DAGJK}.
(2) \textbf{Bootstrap ensemble (``bootstrap/OOB'').}
  For each member $m$, we sample $N$ records \emph{with replacement} from the $N$-record training set. The
  out-of-bag (OOB) records not selected in that bootstrap replicate are used as the validation set for early stopping.

Within each ensemble member, preprocessing (descriptor standardization and univariate descriptor selection) is fit
using only that member's training subset, and then applied to its validation subset and to all inference candidates.

\paragraph{Point predictions and uncertainty summaries.}
For a query condition $x$ (cation SMILES, anion SMILES, and thermodynamic conditions), the ensemble produces
predictions $\{y_m(x)\}_{m=1}^M$ (where $y=\ln(\eta)$ for viscosity and $y=x_{\mathrm{CO_2}}$ for solubility).
We report the ensemble mean as the point estimate,
\begin{equation}
\mu(x) = \frac{1}{M}\sum_{m=1}^M y_m(x),
\end{equation}
and the ensemble standard deviation as an epistemic uncertainty proxy,
\begin{equation}
s(x) = \sqrt{\frac{1}{M-1}\sum_{m=1}^M \left(y_m(x)-\mu(x)\right)^2 }.
\end{equation}
We also compute an empirical central 90\% ensemble interval width,
\begin{equation}
w_{90}(x) = Q_{0.95}\left(\{y_m(x)\}\right) - Q_{0.05}\left(\{y_m(x)\}\right),
\end{equation}
as a more interpretable uncertainty summary for large-scale screening. For viscosity, uncertainties are naturally
interpreted in log-space; an additive uncertainty $\Delta$ in $\ln(\eta)$ corresponds to a multiplicative factor
$\exp(\Delta)$ in $\eta$.

\paragraph{Agreement diagnostics.}
For every model pair we compute pairwise Pearson correlation, RMSE, MAE, mean bias, and the Bland--Altman limits-of-agreement half-width ($1.96$ times the standard deviation of the pairwise prediction differences).

\subsubsection{Uncertainty and model agreement results}
\label{sec:uncertainty}

\paragraph{Epistemic uncertainty and ensemble agreement.}
At the absorption-condition screening point (viscosity at $T=313.15$~K; CO$_2$ solubility at $T=313.15$~K, $P=1$~bar), we summarize the ensemble dispersion $s(x)$ and $90\%$ interval width $w_{90}(x)$ over the full candidate space ($N=408{,}634$) in Table~\ref{tab:uq_summary}. The subsampled (dropout90/val10) ensemble yields mean $\mathbb{E}[s]=0.238$ in $\ln(\eta)$ and $0.023$ in $x_{\mathrm{CO_2}}$, with the bootstrap ensemble systematically larger. Pairwise member agreement is high for the subsampled ensembles ($r=0.958$--$0.971$ for viscosity, $0.921$--$0.951$ for CO$_2$) with near-zero inter-member bias, indicating no single member dominates the dispersion. Across schemes, mean predictions are highly consistent (viscosity $r=0.988$; CO$_2$ $r=0.979$) while uncertainty estimates agree only moderately (std $r=0.535$ and $0.702$); we therefore use the subsampled ensemble as primary and report bootstrap as a robustness check.

\begin{table}[t]
\centering
\tiny
\caption{Summary of ensemble uncertainty distributions over the absorption-condition screening set
($N=408{,}634$ candidate ILs). For viscosity, statistics are in $\ln(\eta)$ units; for CO$_2$ solubility, in
$x_{\mathrm{CO_2}}$ units. $w_{90}$ denotes the empirical ensemble 90\% interval width $Q_{0.95}-Q_{0.05}$.}
\label{tab:uq_summary}
\setlength{\tabcolsep}{4.5pt}
\begin{tabular}{lrrrrrrr}
\toprule
Dataset & $M$ & $N$ & mean $s$ & median $s$ & p95($s$) & mean $w_{90}$ & p95($w_{90}$) \\
\midrule
Viscosity (subsample 90\%) & 5 & 408634 & 0.238 & 0.210 & 0.487 & 0.520 & 1.067 \\
Viscosity (bootstrap)      & 5 & 408634 & 0.272 & 0.235 & 0.570 & 0.595 & 1.246 \\
CO$_2$ (subsample 90\%)    & 5 & 408634 & 0.023 & 0.012 & 0.072 & 0.049 & 0.157 \\
CO$_2$ (bootstrap)         & 5 & 408634 & 0.030 & 0.017 & 0.093 & 0.064 & 0.202 \\
\bottomrule
\end{tabular}
\end{table}

\section{Probabilistic Pareto selection methodology}
\label{app:pareto_objectives}

\paragraph{Ensemble-based predictive worlds.}
We rank IL candidates under predictive uncertainty using an ensemble of $B$ joint predictive ``worlds.'' Each world $b \in \{1, \ldots, B\}$ provides a joint realization of all objectives for every candidate, preserving cross-objective correlations induced by the underlying models. Let $y_i^{(b)} \in \mathbb{R}^M$ denote the $M$-objective vector for candidate $i$ in world $b$.

\paragraph{Derived objectives.}
In each world $b$, working capacity $\Delta C_i^{(b)}$ is computed from the per-world solubility predictions via Eq.~(1) ($T_{\mathrm{abs}}=313.15$~K, $T_{\mathrm{des}}=393.15$~K), and heat of absorption via a Van't Hoff regression:

The viscosity objective is the predicted log-viscosity at $T_{\mathrm{abs}}$:
\begin{equation}
\ln\eta_i^{(b)} = \ln\left(\eta_i^{(b)}(T_{\mathrm{abs}})\right).
\end{equation}

\paragraph{Feasibility screening.}
Before Pareto computations, we remove weak CO$_2$ binders by enforcing:
\begin{equation}
\left|\Delta H_i^{(b)}\right| > H_{\min}, \qquad H_{\min} = 5~\mathrm{kJ/mol}.
\end{equation}
Candidates failing this criterion are excluded from subsequent Pareto computations.

\paragraph{Pareto ranking}
Objectives are converted to a common minimization sense, and per-world dominance and the non-dominance frequency $P_{\mathrm{ndom}}(i)$ follow Eqs.~(3)--(4) of Section~\ref{sec:pareto}, with deterministic tie-breaking as needed. Optionally, epsilon-dominance variants and per-objective summary statistics characterize uncertainty among top-ranked candidates.

%% file: bibliography.bib
@Article{D3DD00040K,
  author    = "Chen, Guzhong and Song, Zhen and Qi, Zhiwen and Sundmacher, Kai",
  title     = "Generalizing property prediction of ionic liquids from limited labeled data: a one-stop framework empowered by transfer learning",
  journal   = "Digital Discovery",
  year      = "2023",
  volume    = "2",
  issue     = "3",
  pages     = "591-601",
  publisher = "RSC",
  doi       = "10.1039/D3DD00040K",
  url       = "http://dx.doi.org/10.1039/D3DD00040K"
}

@Article{data4020088,
  author    = {Venkatraman, Vishwesh and Evjen, Sigvart and Chellappan Lethesh, Kallidanthiyil},
  title     = {The Ionic Liquid Property Explorer: An Extensive Library of Task-Specific Solvents},
  journal   = {Data},
  volume    = {4},
  year      = {2019},
  number    = {2},
  article-number = {88},
  url       = {https://www.mdpi.com/2306-5729/4/2/88},
  issn      = {2306-5729},
  doi       = {10.3390/data4020088}
}

@article{CHEN2025335,
  title     = {Exploring the chemical space of ionic liquids for CO$_2$ dissolution through generative machine learning models},
  journal   = {Green Chemical Engineering},
  volume    = {6},
  number    = {3},
  pages     = {335-343},
  year      = {2025},
  issn      = {2666-9528},
  doi       = {https://doi.org/10.1016/j.gce.2024.06.005},
  url       = {https://www.sciencedirect.com/science/article/pii/S2666952824000414},
  author    = {Xiuxian Chen and Guzhong Chen and Kunchi Xie and Jie Cheng and Jiahui Chen and Zhen Song and Zhiwen Qi}
}

@article{Dong2007,
  author    = {Dong, Q. and Muzny, C. D. and Kazakov, A. and Diky, V. and Magee, J. W. and Widegren, J. A. and Chirico, R. D. and Marsh, K. N. and Frenkel, M.},
  title     = {ILThermo: A Free-Access Web Database for Thermodynamic Properties of Ionic Liquids},
  journal   = {Journal of Chemical \& Engineering Data},
  year      = {2007},
  volume    = {52},
  number    = {4},
  pages     = {1151--1159},
  doi       = {10.1021/je700171f},
  url       = {https://doi.org/10.1021/je700171f}
}

@article{Heid2023,
    author = {Heid, Esther and Greenman, Kevin P. and Chung, Yunsie and Li, Shih-Cheng and Graff, David E. and Vermeire, Florence H. and Wu, Haoyang and Green, William H. and McGill, Charles J.},
    title = {Chemprop: A Machine Learning Package for Chemical Property Prediction},
    journal = {Journal of Chemical Information and Modeling},
    volume = {64},
    number = {1},
    pages = {9-17},
    year = {2024},
    doi = {10.1021/acs.jcim.3c01250},
        note ={PMID: 38147829},
    URL = { 
            https://doi.org/10.1021/acs.jcim.3c01250
    },
    eprint = {     
            https://doi.org/10.1021/acs.jcim.3c01250
    }
}

@article{Raganati2020,
    author = {Raganati, Federica and Chirone, Riccardo and Ammendola, Paola},
    title = {CO$_2$ Capture by Temperature Swing Adsorption: Working Capacity As Affected by Temperature and CO$_2$ Partial Pressure},
    journal = {Industrial \& Engineering Chemistry Research},
    volume = {59},
    number = {8},
    pages = {3593-3605},
    year = {2020},
    doi = {10.1021/acs.iecr.9b04901},
    URL = {https://doi.org/10.1021/acs.iecr.9b04901},
    eprint = {https://doi.org/10.1021/acs.iecr.9b04901}
}

@article{Rochelle2009,
  title={Amine scrubbing for CO$_2$ capture},
  author={Rochelle, Gary T.},
  journal={Science},
  volume={325},
  number={5948},
  pages={1652--1654},
  year={2009},
  publisher={American Association for the Advancement of Science},
  doi={10.1126/science.1176731}
}

@article{Mehrkesh2016,
   title={New quantum chemistry-based descriptors for better prediction of melting point and viscosity of ionic liquids},
   volume={427},
   ISSN={0378-3812},
   url={http://dx.doi.org/10.1016/j.fluid.2016.07.006},
   DOI={10.1016/j.fluid.2016.07.006},
   journal={Fluid Phase Equilibria},
   publisher={Elsevier BV},
   author={Mehrkesh, Amirhossein and Karunanithi, Arunprakash T.},
   year={2016},
   month=nov, pages={498–503} }

@article{Vermeire_2021,
   title={Transfer learning for solvation free energies: From quantum chemistry to experiments},
   volume={418},
   ISSN={1385-8947},
   url={http://dx.doi.org/10.1016/j.cej.2021.129307},
   DOI={10.1016/j.cej.2021.129307},
   journal={Chemical Engineering Journal},
   publisher={Elsevier BV},
   author={Vermeire, Florence H. and Green, William H.},
   year={2021},
   month=aug, pages={129307} }

@article{Racki2025,
    author = {Racki, Adrian and Paduszyński, Kamil},
    title = {Recent Advances in the Modeling of Ionic Liquids Using Artificial Neural Networks},
    journal = {Journal of Chemical Information and Modeling},
    volume = {65},
    number = {7},
    pages = {3161-3175},
    year = {2025},
    doi = {10.1021/acs.jcim.4c02364},
        note ={PMID: 40143756},
    URL = { https://doi.org/10.1021/acs.jcim.4c02364},
    eprint = {https://doi.org/10.1021/acs.jcim.4c02364}
}

@article{Vinjarapu2024,
    title = {Heat of absorption of CO$_2$ in 30 wt% MEA with monoethyleneglycol and urea as vapour reduction additives},
    journal = {Energy},
    volume = {293},
    pages = {130609},
    year = {2024},
    issn = {0360-5442},
    doi = {https://doi.org/10.1016/j.energy.2024.130609},
    url = {https://www.sciencedirect.com/science/article/pii/S0360544224003815},
    author = {Sai Hema Bhavya Vinjarapu and Teresa Regueira and Randi Neerup and Nicolas {von Solms} and Philip Loldrup Fosbøl},
}

@article{Vega2016,
    title = {Study of the MEA degradation in a CO$_2$ capture process based on partial oxy-combustion approach},
    journal = {International Journal of Greenhouse Gas Control},
    volume = {54},
    pages = {160-167},
    year = {2016},
    issn = {1750-5836},
    doi = {https://doi.org/10.1016/j.ijggc.2016.09.007},
    url = {https://www.sciencedirect.com/science/article/pii/S1750583616305886},
    author = {F. Vega and A. Sanna and M.M. Maroto-Valer and B. Navarrete and D. Abad-Correa},
}

@article{Koutsoukos2021ILReview,
  title   = {A review on machine learning algorithms for the ionic liquid chemical space},
  author  = {Koutsoukos, Spyridon and Philippi, Frederik and Malaret, Francisco and Welton, Tom},
  journal = {Chemical Science},
  year    = {2021},
  volume  = {12},
  number  = {20},
  pages   = {6820--6843},
  doi     = {10.1039/D1SC01000J}
}

@article{Chen2022ViscosityILWater,
  title   = {Machine Learning for the Prediction of Viscosity of Ionic Liquid--Water Mixtures},
  author  = {Chen, Y. and Peng, B. and Kontogeorgis, G. and Liang, X.},
  journal = {Journal of Molecular Liquids},
  year    = {2022},
  volume  = {350},
  pages   = {118546},
  doi     = {10.1016/j.molliq.2022.118546}
}

@article{Song2020CO2IL,
  title   = {Prediction of CO$_2$ solubility in ionic liquids using machine learning methods},
  author  = {Song, Zhen and Shi, Huaiwei and Zhang, Xiang and Zhou, Teng},
  journal = {Chemical Engineering Science},
  year    = {2020},
  volume  = {223},
  pages   = {115752},
  doi     = {10.1016/j.ces.2020.115752}
}

@article{Ali2024CO2IL,
  title   = {Prediction of CO$_2$ solubility in Ionic liquids for CO$_2$ capture using deep learning models},
  author  = {Ali, Mazhar and Sarwar, Tooba and Mubarak, Nabisab Mujawar and Karri, Rama Rao and Ghalib, Lubna and Bibi, Aisha and Mazari, Shaukat Ali},
  journal = {Scientific Reports},
  year    = {2024},
  volume  = {14},
  pages   = {14730},
  doi     = {10.1038/s41598-024-65499-y}
}

@article{Laakso2025SigmaProfileCO2,
  title   = {Machine learning modeling of the CO$_2$ solubility in ionic liquids by using $\sigma$-profile descriptors},
  author  = {Laakso, Juho-Pekka and Gorji, Ali Ebrahimpoor and Uusi-Kyyny, Petri and Alopaeus, Ville},
  journal = {Chemical Engineering Science},
  year    = {2025},
  pages   = {121226},
  doi     = {10.1016/j.ces.2025.121226}
}

@inproceedings{Lakshminarayanan2017DeepEnsembles,
  title     = {Simple and Scalable Predictive Uncertainty Estimation using Deep Ensembles},
  author    = {Lakshminarayanan, Balaji and Pritzel, Alexander and Blundell, Charles},
  booktitle = {Advances in Neural Information Processing Systems (NeurIPS)},
  year      = {2017}
}

@inproceedings{Gross1980,
  title        = {Median Estimation in Sample Surveys},
  author       = {Gross, S.},
  year         = {1980},
  booktitle    = {Proceedings of the American Statistical Association, Section on Survey Research Methods},
  pages        = {181--184},
  organization = {American Statistical Association}
}

@article{WangPengKim2022FinitePopBootstrap,
  title   = {Bootstrap inference for the finite population mean under complex sampling designs},
  author  = {Wang, Zhonglei and Peng, Liuhua and Kim, Jae Kwang},
  journal = {Journal of the Royal Statistical Society: Series B (Statistical Methodology)},
  year    = {2022},
  volume  = {84},
  number  = {4},
  pages   = {1150--1174},
  doi     = {10.1111/rssb.12506}
}

@article{Kott2001DAGJK,
  title   = {The Delete-a-Group Jackknife},
  author  = {Kott, Phillip S.},
  journal = {Journal of Official Statistics},
  year    = {2001},
  volume  = {17},
  number  = {4},
  pages   = {521--526}
}

@article{Saigo2010FourBootstraps,
  title   = {Comparing Four Bootstrap Methods for Stratified Three-Stage Sampling},
  author  = {Saigo, Hiroshi},
  journal = {Journal of Official Statistics},
  year    = {2010},
  volume  = {26},
  number  = {1},
  pages   = {193--207}
}

@inproceedings{Daulton2021NEHVI,
  title     = {Parallel Bayesian Optimization of Multiple Noisy Objectives with Expected Hypervolume Improvement},
  author    = {Daulton, Samuel and Balandat, Maximilian and Bakshy, Eytan},
  booktitle = {Advances in Neural Information Processing Systems (NeurIPS)},
  year      = {2021}
}

@article{Jablonka2021BiasFreeMOAL,
  title   = {Bias free multiobjective active learning for materials design and discovery},
  author  = {Jablonka, Kevin Maik and Jothiappan, Giriprasad Melpatti and Wang, Shefang and Smit, Berend and Yoo, Brian},
  journal = {Nature Communications},
  year    = {2021},
  volume  = {12},
  pages   = {2312},
  doi     = {10.1038/s41467-021-22437-0}
}

@article{Low2023MappingPareto,
  title   = {Mapping pareto fronts for efficient multi-objective materials discovery},
  author  = {Low, Andre K. Y. and Vissol-Gaudin, Eleonore and Hippalgaonkar, Kedar},
  journal = {Journal of Materials Informatics},
  year    = {2023},
  volume  = {3},
  pages   = {11},
  doi     = {10.20517/jmi.2023.02}
}

@article{Jiang2023AIRetrosynthesisReview,
  title   = {Artificial Intelligence for Retrosynthesis Prediction},
  author  = {Jiang, Yinjie and Yu, Yemin and Kong, Ming and Mei, Yu and Yuan, Luotian and Huang, Zhengxing and Kuang, Kun and Wang, Zhihua and Yao, Huaxiu and Zou, James and Coley, Connor W. and Wei, Ying},
  journal = {Engineering},
  year    = {2023},
  volume  = {25},
  pages   = {32--50},
  doi     = {10.1016/j.eng.2022.04.021}
}

@article{Ertl2009SAscore,
  title   = {Estimation of synthetic accessibility score of drug-like molecules based on molecular complexity and fragment contributions},
  author  = {Ertl, Peter and Schuffenhauer, Ansgar},
  journal = {Journal of Cheminformatics},
  year    = {2009},
  volume  = {1},
  pages   = {8},
  doi     = {10.1186/1758-2946-1-8}
}

@article{Gao2025RetroScore,
  title   = {RetroScore: graph edit distance-guided retrosynthesis for accessibility scoring with route metrics},
  author  = {Gao, S. and others},
  journal = {Journal of Cheminformatics},
  year    = {2025},
  doi     = {10.1186/s13321-025-01138-6}
}

@article{miao2014chemical,
  title={On the Chemical Stabilities of Ionic Liquids},
  author={Miao, Wang and Ting Hua, Chan and others},
  journal={Molecules},
  volume={19},
  number={2},
  pages={1478--1506},
  year={2014},
  publisher={MDPI}
}

@article{borun2019conductance,
  title={Conductance and ionic association of selected imidazolium ionic liquids in various solvents: A review},
  author={Boru{\'n}, Agnieszka},
  journal={Journal of Molecular Liquids},
  volume={276},
  pages={14--31},
  year={2019},
  publisher={Elsevier}
}

@article{ghandi2014review,
  title={A Review of Ionic Liquids, Their Limits and Applications},
  author={Ghandi, Khashayar},
  journal={Green and Sustainable Chemistry},
  volume={4},
  number={1},
  pages={44--53},
  year={2014},
  publisher={Scientific Research Publishing}
}

@article{VENKATRAMAN2017162,
title = {Predicting CO2 capture of ionic liquids using machine learning},
journal = {Journal of CO2 Utilization},
volume = {21},
pages = {162-168},
year = {2017},
issn = {2212-9820},
doi = {https://doi.org/10.1016/j.jcou.2017.06.012},
url = {https://www.sciencedirect.com/science/article/pii/S2212982016304917},
author = {Vishwesh Venkatraman and Bjørn Kåre Alsberg}
}

@article{SONG2020115752,
title = {Prediction of CO2 solubility in ionic liquids using machine learning methods},
journal = {Chemical Engineering Science},
volume = {223},
pages = {115752},
year = {2020},
issn = {0009-2509},
doi = {https://doi.org/10.1016/j.ces.2020.115752},
url = {https://www.sciencedirect.com/science/article/pii/S0009250920302840},
author = {Zhen Song and Huaiwei Shi and Xiang Zhang and Teng Zhou}
}
